\documentclass[reprint, amsmath,amssymb,prb]{revtex4-2}

\usepackage{color}
\usepackage{graphicx}
\usepackage{dcolumn}
\usepackage{bm}

\begin{document}


\title{Radial Kohn-Sham problem via integral-equation approach}
\author{J\={a}nis U\v{z}ulis}
\author{Andris Gulans}
\affiliation{Department of Physics, University of Latvia, Jelgavas iela 3, Riga, LV-1058 Latvia}

\date{\today}

\begin{abstract}
We present a numerical tool for solving the non-relativistic Kohn-Sham problem for spherically-symmetric atoms.
It treats the Schr\"{o}dinger equation as an integral equation relying heavily on convolutions.
The solver supports different types of exchange-correlation functionals including screened and long-range corrected hybrids. 
We implement a new method for treating range separation based on the complementary error function kernel.
The present tool is applied in non-relativistic total energy calculations of atoms.
A comparison with ultra-precise reference data[Cinal, JOMC 58, 1571 (2020)] shows a 14-digit agreement for Hartree-Fock results.
We provide further benchmark data obtained with 5 different exchange-correlation functionals.
\end{abstract}

\maketitle


\section{\label{sec:introduction}Introduction}

Particle in a spherically symmetric potential is a classic problem that enters quantum mechanics textbooks.
This problem has a fundamental importance in modern computational materials science, as it has to be solved in electronic-structure codes for a variety of purposes.
One comes across it when generating pseudopotentials \cite{Troullier1991},
constructing basis functions in several formalisms such as numerical atomic orbitals\cite{Blum2009}, linearized augmented plane waves (LAPW) and linearized muffin-tin orbitals\cite{Slater1937,Andersen1975}.
This problem appears also in a calculation of core orbitals in LAPW.

The specific form of the spherically symmetric problem that raises interest in the context of electronic-structure calculations is defined by the employed method.
The majority of these calculations rely on (semi)local density-functional theory (DFT), as evident from data-centered high-throughput initiatives \cite{Curtarolo2012,Jain2013,Kirklin2015}. 
In this case, the exchange and correlation are described by means of the local-density approximation (LDA) or the generalized gradient approximation (GGA).
These models lead to the Kohn-Sham (KS) equation 
\begin{equation}\label{eq:rselocal}
-\frac{\nabla^2}{2}\psi(\mathbf{r})+v(\mathbf{r})\psi(\mathbf{r})=\varepsilon\psi(\mathbf{r})
\end{equation}
with a local effective KS potential $v(\mathbf{r})$.
Assuming the spherical symmetry of the potential, this equation reduces to a one-dimensional problem that can be solved by an outward integration as an initial value problem.
This task is typically performed using the Numerov's method\cite{Numerov1924}, although other options such as the predictor-corrector and Bulirsch-Stoer\cite{Bulirsch1966} method are also applied in electronic-structure codes~\cite{elk,Gulans2014}. 
If Eq.~\ref{eq:rselocal} is intended as an eigenvalue problem, the outward integration is coupled with the shooting method.
In other words, the procedure of solving the radial problem is already well-established in the case of a local potential.

The (semi)local approximations in DFT perform with limitations for a range of applications that require accurate atomization (reaction) energies \cite{Peverati2014} or good estimates of the band structure.
A frequently applied solution to this problem is hybrid exchange-correlation functionals where the GGA and/or LDA are combined with the Fock, screened \cite{Heyd2003} or the long-range \cite{Yoshihiro2004} exchange.
These three exchange models introduce a non-local potential $\hat{v}^\mathrm{nl}$ that makes the usual approach to solving the radial equation not applicable. 
Two recent studies addressed this problem with an aim to implement an efficient and precise solver \cite{Cinal2020,Lehtola2020}.
Cinal solved the Hartree-Fock (HF) equations using the pseudospectral method \cite{Cinal2020}, but did not consider DFT.
The obtained atomic energies were computed in double and quadruple precision ensuring an extreme level of precision. 
He employed compact a Lobatto-type grid \cite{abramowitz1965handbook} that is not consistent with the codes mentioned above.

In the other study, Lehtola implemented an atomic solver based on the finite-element method~\cite{Lehtola2019,Lehtola2020} employing a high-order basis with a small number of support points.
He applied the code in HF and DFT calculations and reported total energies in Ha with six decimals for all elements up to atomic number $118$.
His study relied on a generalized exponential support grid, but the method can be used with any type of a grid~\cite{Lehtola2019}.
Ref.~\cite{Lehtola2020} also presents an implementation of hybrid functionals with range separation and uses it for providing benchmark energies for light spherically-symmetric atoms and ions.
The implementation is based on a spherical harmonic expansion of the complementary error function (erfc) kernel leading to a bi-variate quadrature as introduced in Ref.~\cite{Angyan2006}. 
So-obtained atomic energies calculated employing the LC-BLYP functional \cite{Yoshihiro2004} were verified using only a Gaussian basis-set calculation with a limited precision, since there were no other data for a comparison.

In this study, we propose an alternative precise approach for solving the spherically-symmetric problem and therefore provide an independent high-quality reference for atomic calculations. 
It is suitable for calculations with the types of grids that are employed in electronic-structure codes and can be integrated into them.
Our method uses the idea that the Schr\"{o}dinger equation can be rewritten as an integral equation following the ideas first published by Kalos in 1962\cite{Kalos1962}.
In literature, this method is known as Helmholtz kernel~\cite{Solala2017}, Green iteration~\cite{Vaughn2021} and Lippmann-Schwinger \cite{Harrison2004} approaches.
It is commonly used in calculations with wavelet basis sets in Ref.~\cite{Harrison2004, Bischoff2011, Bischoff2012, Jensen2017, Ratcliff2020}, because it allows one to avoid computation of derivatives.  
This method was also applied in problems with model potentials \cite{Hu2000}, for optimising orbitals in small molecules within the HF theory \cite{Solala2017} and for octree-based real-space all-electron DFT computations \cite{Vaughn2021}.
Despite this list of applications, we are not aware of previous studies using this method specifically in radial solvers.

We also address the treatment of the erfc kernel in functionals with the screened or range-separated exchange. 
It does not cause any major difficulties in plane-wave and Gaussian-basis calculations due to existence of simple analytical expressions for Fourier transforms and electron-repulsion integrals. In contrast, electronic-structure codes employing LAPWs and Slater orbitals do not benefit from this advantage immediately, and a common workaround involves abandoning erfc and employing the simpler Yukawa kernel implemented in these formalisms~\cite{Tran2011,Rico2012}.
Another workaround in LAPW requires a use of a product basis\cite{Schlipf2011,Vona2021}, and the precision of such an approach requires a verification.
Also radial calculations require a somewhat involved approach, as the standard technique involves a bi-variate quadrature \cite{Angyan2006}.
We propose a simple alternative where the erfc kernel is expanded in terms of complex Yukawa potentials which makes it applicable in other formalisms the atomic solver presented in this work. 
 
This paper is structured as follows. 
We give an introduction to the integral-equation method in Sec.~\ref{sec:integral} and provide expressions for calculating the total energy in Sec.~\ref{sec:toten}.
The formalism described in these sections relies heavily on convolutions and radial integrals discussed in Secs.~\ref{sec:kernels} and \ref{sec:integrals}, respectively.
The considered types of convolutions involve the Coulomb and Yukawa kernels as well as the erfc kernel. 
Our implementation employs a large number of integral evaluations requiring attention to how they are performed.
We show an efficient high-order approach for computing the integrals on arbitrary grids.
In Sec.~\ref{sec:optimal}, we evaluate the performance of logarithmic and inverse polynomial radial grids.
Finally, total-energy calculations of closed-shell atoms are presented and analysed in Sec.~\ref{sec:results}.

\section{\label{sec:integral}Integral-equation method}

We consider the non-relativistic KS equation shown in Eq.~\ref{eq:rselocal}.
We assume that the potential consists of local and non-local contributions $\hat{v}=v^\mathrm{L}(\mathbf{r})+\hat{v}^\mathrm{NL}$. 
The local part comprises the electron-nuclear, Hartree and the LDA or GGA exchange-correlation terms expressed as 
\begin{equation}
\label{eq:nucl}
    v_\mathrm{n}(\mathbf{r})=-Z/r
\end{equation}
with the nuclear charge $Z$,
\begin{equation}
\label{eq:hartree}
    v_\mathrm{H}(\mathbf{r})=\int\frac{\rho(\mathbf{r}^\prime)}{|\mathbf{r}^\prime-\mathbf{r}|}d \mathbf{r}^\prime
\end{equation}
with the electron density $\rho(\mathbf{r})$,
and
\begin{equation}
\label{eq:vxcl}
 v^\mathrm{L}_\mathrm{xc}(\mathbf{r})=(1-\alpha)v^\mathrm{GGA}_\mathrm{x}(\mathbf{r}) + v^\mathrm{GGA}_\mathrm{c}(\mathbf{r}),
\end{equation} 
respectively.
The non-local part is defined by its action on a trial wavefunction $\psi_n(\mathbf{r})$:
\begin{equation}
\label{eq:vxnl}
    \hat{v}^\mathrm{NL}_\mathrm{x} \psi_n(\mathbf{r}) = \alpha\int v^\mathrm{NL}_\mathrm{x}(\mathbf{r},\mathbf{r}^\prime)\psi_n(\mathbf{r}^\prime)  d \mathbf{r}^\prime
\end{equation}
containing the (screened) Fock exchange
\begin{equation}
\label{eq:vnl}
    v^\mathrm{NL}_\mathrm{x}(\mathbf{r},\mathbf{r}^\prime)=\sum\limits_n f_n \psi_n(\mathbf{r}) V(|\mathbf{r}^\prime-\mathbf{r}|)\psi^\ast_n(\mathbf{r}^\prime),
\end{equation}
where the sum runs over all considered orbitals, $f_n$ is the occupation number and $V(|\mathbf{r}|)$ is the interaction kernel (Coloumb, Yukawa, erfc or erf).
Finally, the parameter $\alpha$ is the weight of the non-local exchange in the potential.
We evaluate $v^\mathrm{L}_\mathrm{xc}(\mathbf{r})$ using the \texttt{libxc} library\cite{Lehtola2018_libxc}.
The details on calculating $v_\mathrm{H}(\mathbf{r})$ and $\hat{v}^\mathrm{NL}_\mathrm{x} \psi_n(\mathbf{r})$ are given in Sec.~\ref{sec:kernels}.

The effective potential depends on the electron density, and, therefore, the KS problem has to be solved self-consistently. 
There are numerous methods for ensuring convergence to the ground-state solution (see the review in Ref.~\cite{Woods2019}), but a method as straightforward as linear mixing of potentials works sufficiently well for spherical atoms. 
Therefore, we focus on solving the KS equation for a fixed potential. 

A formal rearrangement of terms in Eq.~\ref{eq:rselocal} leads to 
\begin{equation}\label{eq:formal}
    \psi(\mathbf{r})=2(\nabla^2 -\lambda^2)^ {-1}[\hat{v}\psi(\mathbf{r})]
\end{equation}
with $\lambda^2=-2\varepsilon$.
An expression with $\psi(\mathbf{r})=(\nabla^2 -\lambda^2)^{-1} f(\mathbf{r})$ has the meaning that $\psi(\mathbf{r})$ is the solution of the screened Poisson equation
\begin{equation}\label{eq:scrPoisson}
 (\nabla^2 -\lambda^2)\psi(\mathbf{r}) =-f(\mathbf{r}),
\end{equation}
where the usual prefactor $4\pi$ on the right-hand side is omitted.
If this equation is applied in the context of electrostatics, the functions $f(\mathbf{r})$ and $\psi(\mathbf{r})$ have the meaning of the charge and the resulting screened potential, respectively.

We solve the KS equation for bound states meaning that $\varepsilon<0$ and $\psi(\mathbf{r})$ decays to 0 as $r\rightarrow\infty$. 
Then, it is appropriate to express $(\nabla^2 -\lambda^2)^{-1}$ in Eq.~\ref{eq:formal} via its Green's function in the following manner:
\begin{equation}\label{eq:intse}
    \psi(\mathbf{r})=2\int  \frac{e^{-\lambda |\mathbf{r}-\mathbf{r}^\prime|}}{4\pi|\mathbf{r}-\mathbf{r}^\prime|}\hat{v}\psi(\mathbf{r}^\prime) d\mathbf{r}^\prime.
\end{equation}
Thus, the KS problem given in Eq.~\ref{eq:rselocal} is transformed into the integral equation that does not contain any differential operators.
In spherically symmetric atoms, a wavefunction can be expressed according to its quantum numbers $n$, $\ell$ and $m$ as
\begin{equation}
    \psi_{n\ell m}(\mathbf{r})=u_{n\ell}(r)Y_{\ell m}(\hat{r}),
\end{equation}
where $u_{n\ell}(r)$ is the radial part, and only one spherical harmonic $Y_{\ell m}(\hat{r})$ enters the expression.
A calculation of the convolution in Eq.~\ref{eq:intse} reduces to an evaluation of a one-dimensional integral as discussed below in Sec.~\ref{sec:kernels}. 

Following Ref.~\cite{Kalos1962}, Eq.~\ref{eq:intse} is solved iteratively by inserting a trial function $\psi_{n\ell m}^{(i)}(\mathbf{r})$ to the right-hand side and obtain an updated function $\psi_{n\ell m}^{(i+1)}(\mathbf{r})$, where $i$ is the step number.
The parameter $\lambda$ is updated along with the wavefunction by evaluating 
\begin{equation}
\varepsilon^{(i)}=\langle \psi_{n\ell m}^{(i)} | \hat{H} | \psi_{n\ell m}^{(i)} \rangle / \langle \psi_{n\ell m}^{(i)} | \psi_{n\ell m}^{(i)} \rangle.
\end{equation}
The described procedure converges to the lowest-energy solution for given quantum numbers $\ell$ and $m$.

Now follows a description of a procedure that we use for solving for a few orbitals with a given $\ell$ and $m$ with the lowest Kohn-Sham energies. 
Suppose that, after $i$ iterations, estimates for the KS energies $\varepsilon^{(i)}_{n\ell}$ and the orbitals $\psi_{n\ell m}^{(i)}(\mathbf{r})$ have been obtained.
In step 1, we construct basis functions 
\begin{equation}
\label{eq:step1}
\chi_n^{(i)}(\mathbf{r})=2(\nabla^2+2\varepsilon_{n\ell}^{(i)})^{-1}[\hat{v}\psi_{n\ell m}^{(0)}(\mathbf{r})],
\end{equation}
where the initial guess for the wavefunction $\psi_{n\ell m}^{(0)}(\mathbf{r})$ remains without updates during this iterative process.
In step 2, we calculate the matrix elements
\begin{equation}
H_{nn^\prime} = \langle \chi_n^{(i)} | -\nabla^2/2+\hat{v} | \chi_{n^\prime}^{(i)} \rangle
\end{equation}
and
\begin{equation}
S_{nn^\prime} = \langle \chi_n^{(i)} |  \chi_{n^\prime}^{(i)} \rangle.
\end{equation}
In step 3, we solve the matrix eigenproblem
\begin{equation}
\label{eq:eigen}
Hz=\sigma Sz,
\end{equation}
where $\sigma$ and $z$ are an eigenvalue and an eigenvector, respectively.
In step 4, we update the estimates 
\begin{equation}
\varepsilon_{n\ell}^{(i+1)}=\sigma_n
\end{equation}
and
\begin{equation}
\label{eq:step4}
\psi_{n\ell m}^{(i+1)}(\mathbf{r})= \sum_{n^\prime} z_{nn^\prime}\chi_{n^\prime}(\mathbf{r}),
\end{equation}
respectively.
If $|\varepsilon_{n\ell}^{(i+1)}-\varepsilon_{n\ell}^{(i)}|$ is greater than a predefined threshold value, this sequence is repeated from step 1.
Once this process converges, $\varepsilon_{n\ell}^{(i)}$ and $\psi_{n\ell m}^{(i)}$ are estimates for the eigenpairs of the KS equation for a given potential.

\begin{figure}[h]
\includegraphics [width=240pt]{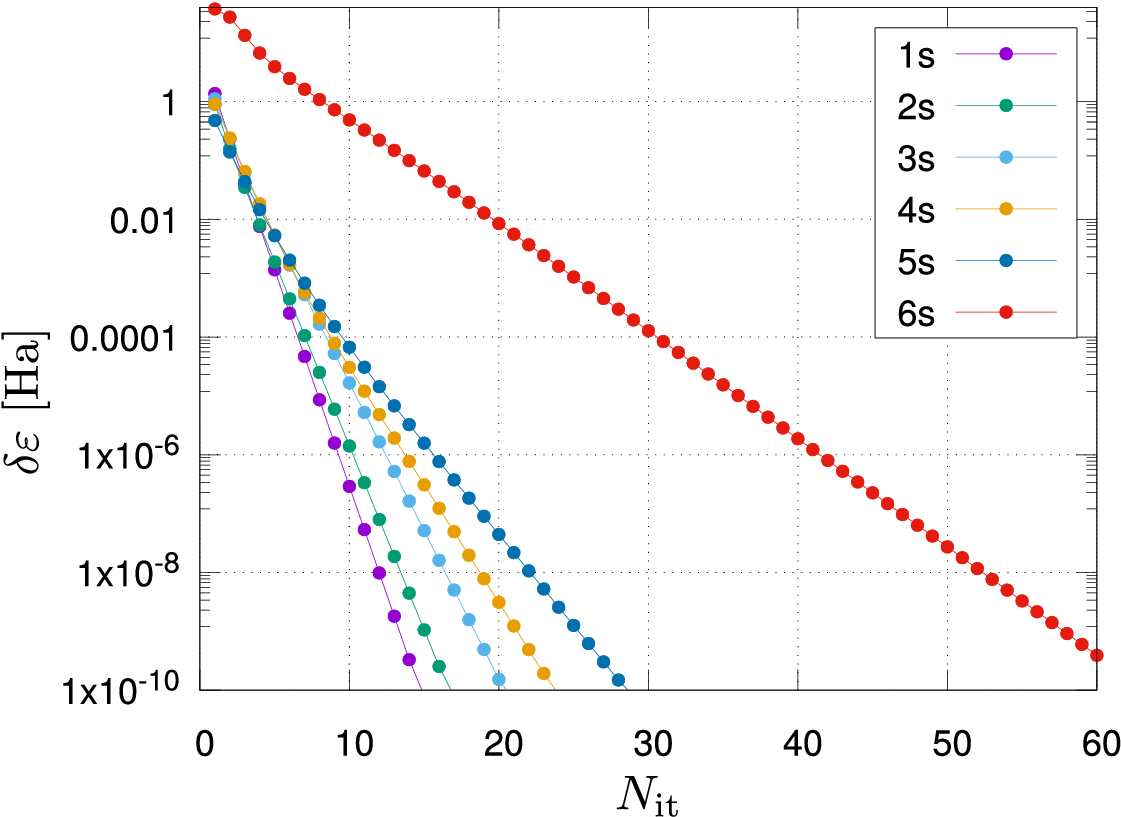}
\caption{\label{fig:LS_conv} Convergence of inner loop (difference between each eigenvalue and its converged value depending on iteration number) for Rn ($\ell=0$ channel) in first external cycle iteration.}
\end{figure}

The iterative procedure introduced in this section differs from what is described in literature. 
Firstly, we fix the input wavefunction on the right-hand-side of Eq.~\ref{eq:step1} to $\psi_{n\ell m}^{(0)}(\mathbf{r})$ as opposed to updating it using $\psi_{n\ell m}^{(i)}(\mathbf{r})$.
Following the latter option, the procedure converges to eigenpairs, whereas our approach generally leads to an approximate result, and its quality depends on the initial guess $\psi_{n\ell m}^{(0)}(\mathbf{r})$.
The procedure defined by Eqs.~\ref{eq:step1}--\ref{eq:step4} is applied multiple times within the self-consistency cycle (every time for a different potential), and, according to our observations, it always converges to the solution of the KS problem with an excellent precision.
The only exception is when KS energies of an atom are close to 0 (typically $\sim -10^{-2}$~Ha).
In such a case, however, the version of the procedure with the update of the wavefunction fails to converge too.

Secondly, we perform a diagonalization (see Eq.~\ref{eq:eigen}) instead of an orthonormalization.
The latter allows Refs.~\cite{Harrison2004} and \cite{Vaughn2021} to avoid a calculation of derivatives. 
It is an important feature for codes employing wavelets and adaptively refined real-space grids in applications beyond atoms.
However, relying on derivatives in the radial problem for spherically-symmetric atoms does not lead to a substantial numerical noise, as evidenced by calculations described in Sec.~\ref{sec:results}. 

To illustrate the performance of the method, we solve the KS equation for the Rn atom. 
The initial guess $\psi_{n\ell m}^{(0)}(\mathbf{r})$ is set to the wavefunctions in hydrogen-like ion with the nuclear charge of Rn.
Fig.~\ref{fig:LS_conv} shows the performance of the described algorithm for the $s$-orbitals.
The KS energies converge to the eigenvalue estimates within $10^{-8}$~Ha in 12--22 and 52 steps in the case of the core and valence orbitals, respectively.
The quality of the initial guess improves with every step of the self-consistency procedure, and the number of the integral-equation iterations reduces dramatically. 

\section{Total energy  \label{sec:toten}}
To calculate the total energy, we consider it as a sum of the five components:
\begin{equation}
    E_\mathrm{tot}=E_\mathrm{kin}+E_\mathrm{n}+E_\mathrm{H}+E_\mathrm{xc}^\mathrm{L}+E_\mathrm{x}^\mathrm{NL}.
\end{equation}
where $E_\mathrm{kin}$ is the kinetic energy and the remaining terms are labelled consistently with the contribution to the effective potential in Eqs.~\ref{eq:nucl}--\ref{eq:vxnl}.
These components are evaluated using radial integrals as follows:
\begin{equation}
E_\mathrm{kin}=
\frac{1}{2}\sum_{n\ell}f_{n\ell} \int\limits_0^\infty \biggl((ru^\prime_{n\ell}(r))^{2} +\ell(\ell+1)u^2_{n\ell}(r)\biggr)  dr,
\end{equation}
where $f_{n\ell}$ is the occupation number of the electron shells,
\begin{equation}
E_\mathrm{ext}=\int_0^\infty -Z \rho(r) r dr,
\end{equation}
\begin{equation}
E_\mathrm{H}=\frac{1}{2}\int_0^\infty v_\mathrm{H}(r) \rho(r) r^2 dr,
\end{equation}
\begin{equation}
E_\mathrm{xc}^L=\int_0^\infty \epsilon_\mathrm{xc}^L(r) \rho(r) r^2 dr,
\end{equation}
where $\epsilon_\mathrm{xc}^L(r)$ is the (semi)local exchange-correlation energy density per particle, and
\begin{equation}
E_\mathrm{x}^\mathrm{NL}=\frac{1}{2}\sum_{n\ell}f_{n\ell} \int_0^\infty  \psi_{n\ell}(r)  [\hat{v}_\mathrm{x}^\mathrm{NL}\psi_{n\ell}(r)] r^2 dr.
\end{equation}

\section{\label{sec:kernels} Convolutions with bare and screened Coulomb kernels}
The algorithm described above employs convolutions with types of kernels:   
(i) the Coulomb kernel $V_\mathrm{C}(\mathbf{r})=1/r$, (ii) the Yukawa kernel $V_\mathrm{Y}(\mathbf{r})=e^{-\lambda r}/r$ and (iii) the erfc kernel $V_\mathrm{SR}(\mathbf{r})=\mathrm{erfc}(\mu r)/r$.

First, we consider the Coulomb kernel $V(\mathbf{r})=1/r$.
It is applied in a calculation of the Hartree potential $v_\mathrm{H}(\mathbf{r})$ in Eq.~\ref{eq:hartree} and the non-local Fock exchange. 
The standard approach is to apply the Laplace expansion:
\begin{equation}{\label{eq:CoublombExpansion}}
\frac{1}{|\mathbf{r}-\mathbf{r}^{\prime}|}=\sum_{\ell=0}^{\infty}\frac{4\pi}{2\ell+1}\frac{r^\ell_<}{r^{\ell+1}_>}Y^\ast_{\ell m} (\hat{r})Y_{\ell m} (\hat{r}^\prime),
\end{equation}
where $r_<=\mathrm{min}(r,r^\prime)$ and $r_>=\mathrm{max}(r,r^\prime)$.
This equation allows us to express the result of the convolution or simply the potential $v(\mathbf{r})$ due to the density $\rho_{\ell m}(r)Y_{\ell m}(\hat{r})$ as 
\begin{eqnarray}
{\label{eq:CoulConv}}
v(\mathbf{r})=\frac{4\pi Y_{\ell m}(\hat{r})}{2\ell+1} 
&\left[ \frac{1}{r^{\ell+1}} \int\limits_0^r r^{\prime\ell+2} \rho_{\ell m}(r^\prime) dr^\prime \right. \nonumber\\
&+ \left.
r^{\ell} \int\limits_r^\infty \frac{1}{r^{\prime\ell-1}} \rho_{\ell m}(r^\prime) dr^\prime \right].
\end{eqnarray}
This equation decouples $r$ from $r^\prime$ and, thus, makes the evaluation of the potential efficient.

In a calculation of the Hartree potential, we assume spherically symmetric density $\rho(\mathbf{r})=\rho_{00}(r) Y_{00}(\hat{r})$.
Also in the case of the Fock exchange, the spherical symmetry is assumed, and the result of the non-local operator acting on a trial wavefunction can be expressed as $\hat{v}_x^\mathrm{NL} \chi_{n\ell m}(\mathbf{r})= \hat{v}_x^\mathrm{NL}\left[ \tilde{u}_{n\ell}(r) Y_{\ell m}(\hat{r}) \right] =h(r) Y_{\ell m}(\hat{r})$.  
Using Eqs.~\ref{eq:vxnl}, \ref{eq:vnl} and \ref{eq:CoublombExpansion}, the radial part of the result is obtained as 
\begin{eqnarray}\label{eq:Fock_all}
h(r) =
&&-\frac{1}{2}\sum_{n^\prime \ell^\prime } \sum\limits_{\ell''=|\ell-\ell'|}^{\ell+\ell'}{\!\!\!\!\!\!\!}{'}
f_{n'\ell'} 
\begin{pmatrix}
\ell & \ell' & \ell''\\
0 & 0 & 0
\end{pmatrix}^2
u_{n'\ell'}(r) \nonumber\\
&&\cdot
\left(\int_0^\infty dr'\frac{r_{<}^{\ell^{''}}}{r_{<}^{\ell^{''}+1}}r'^2 u_{n'\ell'}(r')\tilde{u}_{n\ell}(r')\right),
\end{eqnarray}
where the Wigner 3-j symbol is applied and $\sum^\prime$ has the meaning of of the summations with step 2 (see Ref. \cite{Cinal2020,Johnson2007}).

Similarly to $V_\mathrm{C}(\mathbf{r}-\mathbf{r}^\prime)$, the Yukawa kernel can be written in a separable form:
\begin{equation}
\label{eq:YukawaExpansion}
\frac{e^{-\lambda|\mathbf{r}-\mathbf{r}^{\prime}|}}{|\mathbf{r}-\mathbf{r}^{\prime}|}=
4\pi \lambda \sum_{\ell=0}^{\infty} i_\ell(\lambda r_<) k_\ell(\lambda r_>)\sum_{m=-\ell}^{\ell} Y^*_{\ell m} (\hat{r})Y_{\ell^\prime m^\prime} (\hat{r}^\prime),
\end{equation}
where $\lambda$ is the screening parameter, $i_{\ell}$ and $k_{\ell}$ are the modified spherical Bessel functions of the first and second kinds, respectively.
This equation allows us to factorize the kernel into parts that depend on either on $r$ or $r^\prime$ similarly to how it is done in the case of the Coulomb kernel in Eq.~\ref{eq:CoulConv}.

The third considered convolution kernel $V_\mathrm{SR}(\mathbf{r})$ appears in hybrid exchange-correlation functionals where the range separation is introduced.
These methods decompose the Coulomb interaction into the short- and long-range contributions (SR and LR, respectively) as follows: 
\begin{equation}\label{eq:erfc_split}
\frac{1}{|\mathbf{r}- \mathbf{r}'|}=\underbrace{\frac{\mathrm{erfc}(\mu|\mathbf{r}- \mathbf{r}'|)}{|\mathbf{r}- \mathbf{r}'|}}_{SR}+\underbrace{\frac{\mathrm{erf}(\mu|\mathbf{r}- \mathbf{r}'|)}{|\mathbf{r}- \mathbf{r}'|}}_{LR},
\end{equation}
where $\mu$ is known as the screening or range-separation parameter.
Only one of these two terms remains in the non-local exchange in Eq.~\ref{eq:vnl}, whereas the other one is fully omitted.
Hybrid functionals designed for solid-state applications keep only the SR part and thus mimic screening in bulk materials.  
In calculations of molecules, it is more common to require the correct long-range behaviour of the exchange functional and then the LR term is kept. 

\begin{table}
\caption{Fitting parameters $A_i$ and $a_i$ of the complementary error function following the definition in Eq.~\ref{eq:erfcfit}. Each section of the table begins with a line containing $N_\mathrm{f}$, and $2N_\mathrm{f}+1$ is the total number of fitting functions. 
This line contains also the real part of the exponent $a_i$ which is kept fixed for all $i$.
\label{tab:erfcA}}
\begin{tabular}{rrrr}
\hline
\multicolumn{1}{c}{$i$} & \multicolumn{1}{c}{$\Re(A_i)$} & \multicolumn{1}{c}{$\Im(A_i)$} & \multicolumn{1}{c}{$\Im(a_i)$}\\
\hline
\multicolumn{4}{l}{$N_\mathrm{f}=3$~~~$\Re(a_i)$=3.62435558273902}\\
0 & 3.34075293484632 & 0.00000000000000 & 0.00000000000000 \\
1 & -1.14940001089277 & 1.50422252025891 & 1.43494484832716 \\
2 & -0.02638283853366 & -0.32974525523286 & 2.93170108446798 \\
3 & 0.00540643159457 & 0.01222658029598 & 4.59735152907432 \\
\hline
\multicolumn{4}{l}{$N_\mathrm{f}=4$~~~$\Re(a_i)$=4.03651233059082}\\
0 & 5.88649164984020 & 0.00000000000000 & 0.00000000000000 \\
1 & -2.42999364511904 & 2.89879784656921 & 1.27567126568312 \\
2 & -0.06523181639778 & -0.98148171395886 & 2.58394634761278 \\
3 & 0.05351129281419 & 0.07580726842369 & 3.97319627725743 \\
4 & -0.00153165601674 & -0.00161854017762 & 5.52839978282953 \\
\hline
\multicolumn{4}{l}{$N_\mathrm{f}=5$~~~$\Re(a_i)$=4.44903651057752}\\
0 & 11.68016225012443 & 0.00000000000000 & 0.00000000000000 \\
1 & -5.58814846386175 & 5.91087741873711 & 1.16136522308687 \\
2 & 0.04005009651638 & -2.71301646794361 & 2.34291495991632 \\
3 & 0.22742059158799 & 0.34113912678865 & 3.57030720633364 \\
4 & -0.01968628999525 & -0.01379539810330 & 4.88590055700420 \\
5 & 0.00028294069148 & 0.00022566427890 & 6.37901581267627 \\
\hline
\multicolumn{4}{l}{$N_\mathrm{f}=6$~~~$\Re(a_i)$=4.42897206653722}\\
0 & 10.43888617723020 & 0.00000000000000 & 0.00000000000000 \\
1 & -4.15230200875813 & 6.36698813144879 & 1.07811959900154 \\
2 & -1.09637578338379 & -2.73158403646281 & 2.16055514404309 \\
3 & 0.58534033454693 & 0.16941238316725 & 3.25051246017243 \\
4 & -0.05722146054555 & 0.03586257692680 & 4.35355433640684 \\
5 & 0.00111445947001 & -0.00356355734717 & 5.50798572653769 \\
6 & 0.00000137005560 & 0.00007066173359 & 6.80680166786776 \\
\hline
\multicolumn{4}{l}{$N_\mathrm{f}=7$~~~$\Re(a_i)$=4.56174027039906}\\
0 & 12.86098703034591 & 0.00000000000000 & 0.00000000000000 \\
1 & -4.90510587402478 & 8.42049068837062 & 1.01602163523968 \\
2 & -1.93579633844306 & -3.74443704209627 & 2.04133345120767 \\
3 & 1.01179259239745 & 0.15201429359842 & 3.08051399458118 \\
4 & -0.09901358250763 & 0.09599744431866 & 4.11644818289646 \\
5 & -0.00338136152879 & -0.01174395914925 & 5.06467929848477 \\
6 & 0.00103305167262 & 0.00015238534910 & 5.97856237235781 \\
7 & -0.00002200273877 & 0.00000584510008 & 7.20275534500376 \\
\hline
\multicolumn{4}{l}{$N_\mathrm{f}=8$~~~$\Re(a_i)$=4.55393944225591}\\
0 & 11.79355692851662 & 0.00000000000000 & 0.00000000000000 \\
1 & -3.39975426063274 & 8.58842130087254 & 0.94528885879869 \\
2 & -3.19998047591390 & -3.10247031448467 & 1.89068584187872 \\
3 & 1.24319452630443 & -0.50090970998006 & 2.83585047371902 \\
4 & -0.01592303613588 & 0.25412899494655 & 3.78108079372661 \\
5 & -0.02633000853185 & -0.01569161687470 & 4.72860249427096 \\
6 & 0.00201070487884 & -0.00122269463134 & 5.67051703001310 \\
7 & 0.00000598719210 & 0.00012240497422 & 6.61841395394285 \\
8 & -0.00000190141931 & -0.00000217573991 & 7.70094251177546 \\
\hline
\end{tabular}
\end{table}

As the LR term can be expressed as $V_\mathrm{LR}(\mathbf{r})=V_\mathrm{C}(\mathbf{r})-V_\mathrm{SR}(\mathbf{r})$, we discuss in detail only the SR term (the erfc kernel). 
As shown in previous studies, $V_\mathrm{SR}(\mathbf{r}- \mathbf{r}')$ is not separable in the same way as the Coulomb and Yukawa kernels, and the standard approach to performing such a convolution involves a two-dimensional quadrature~\cite{Angyan2006,Lehtola2020}. 
In this work, we propose an alternative method that reduces the calculation to evaluating a few one-dimensional integrals.
The $\mathrm{erfc}$ function can be efficiently represented in the following form:
\begin{equation}
\label{eq:erfcfit}
\mathrm{erfc}(r) \approx \sum_{j=-N_\mathrm{f}}^{N_\mathrm{f}} A_j e^{a_j r},
\end{equation}
where $A_j$ and $a_j$ are complex parameters determined in a fitting procedure.
To ensure that $\mathrm{erfc}(r)$ is a real function, we impose that $A_j=A^\ast_{-j}$, $a_j=a^\ast_{-j}$ as well as $A_0$ and $a_0$ are real. 
The obtained fitting parameters are given Tab.~\ref{tab:erfcA}.

To justify the approximation in Eq.~\ref{eq:erfcfit}, we consider the function $F(t)=\mathrm{erfc}(r/2\sqrt{t})$.
Following Ref.~\cite{Marshall2002}, its Laplace transform reads
\begin{equation}
\tilde{F}(s)=\frac{e^{-\sqrt{s}r}}{s}.
\end{equation}
Using the Bromwich integral for the inverse Laplace transform, we express $F(t)$ as follows:
\begin{equation}
\mathrm{erfc}(r/2\sqrt{t})=\frac{1}{2\pi i}\lim_{T\to+\infty}\int\limits_{\gamma-iT}^{\gamma+iT} \frac{e^{st} e^{-\sqrt{s}r}}{s} ds,
\end{equation}
where $\gamma$ is a real number. 
Since the integrand has no singularities, $\gamma$ can be chosen freely.
We approximate the expression by a quadrature and obtain
\begin{equation}
\label{eq:LTquad}
\mathrm{erfc}(r/2\sqrt{t})\approx \sum\limits_{j=-N_\mathrm{f}}^{N_\mathrm{f}} W_j \frac{e^{s_j t} e^{-\sqrt{s_j}r}}{s_j},
\end{equation}
where $W_j$ are quadrature weights with the prefactor $1/(2\pi i)$ absorbed, and $s_j$ are complex quadrature grid points.
Comparing Eqs.~\ref{eq:erfcfit} and \ref{eq:LTquad}, we recognize that both equations have the same structure with $t=1/4$ and the complex parameters expressed as $A_j=W_j e^{s_j/4}/s_j$ and $a_j=\sqrt{s_j}$.
Thus, Eq.~\ref{eq:erfcfit} has a clear mathematical interpretation, i.e., it is an identity approximated via a quadrature. 

\begin{figure}[t]
\includegraphics [width=240pt]{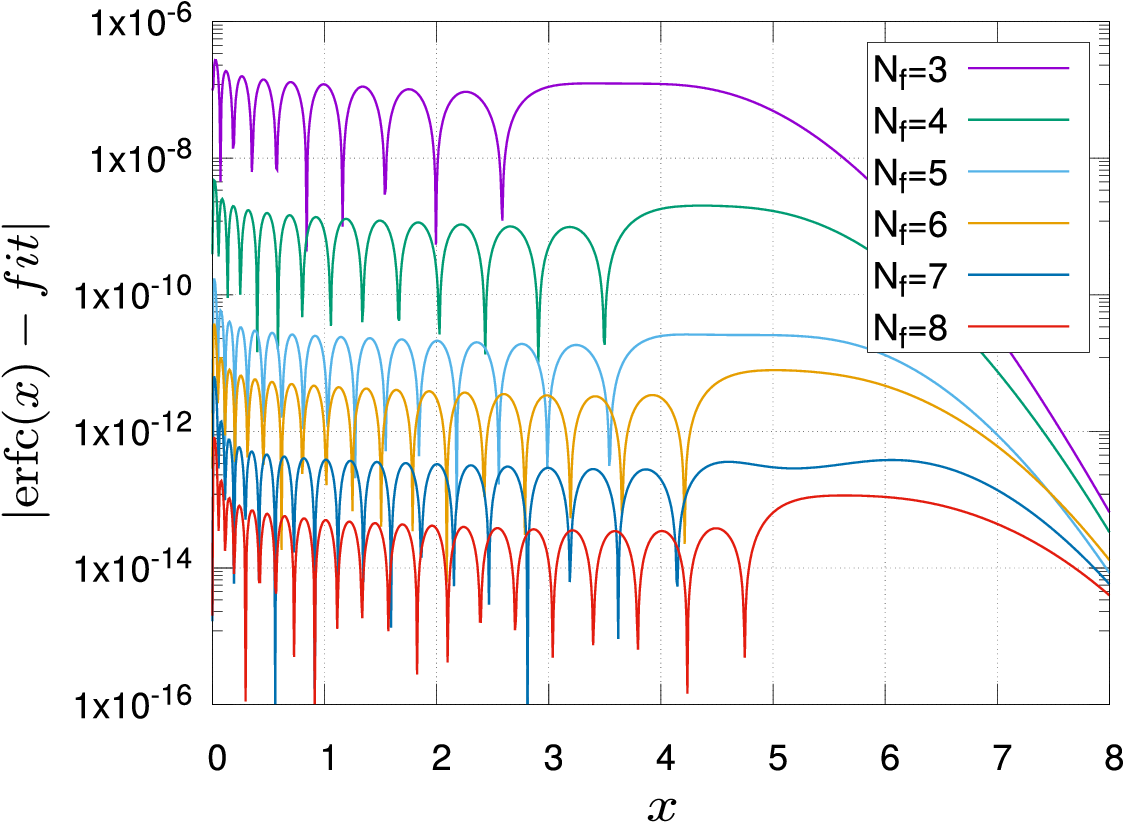}
\caption{\label{fig:erfc_fit} Unsigned difference between complementary error function ($\mathrm{erfc}$) and its fit with $2N_f+1$ functions in Eq.~\ref{eq:erfcfit}. }
\end{figure}

Fig.~\ref{fig:erfc_fit} shows the quality of the fit depending on the number of the functions.
Increasing the number of the exponents in Eq.~\ref{eq:erfcfit} reduces the error of the representation. 
A linear combination of 17 functions ($N_\mathrm{f}=8$) reproduces the $\mathrm{erfc}(r)$ values within $\sim 10^{-12}$ for all allowed values of $r$ and within $\sim 10^{-13}$ for all $r>0.1$.

Using the fit, we express $V_\mathrm{SR}(\mathbf{r}-\mathbf{r}^\prime)$ as a linear combination of complex Yukawa potentials:
\begin{equation}
\label{eq:sry}
V_\mathrm{SR}(\mathbf{r}-\mathbf{r}^\prime)=\sum_{j=-N_{f}}^{N_{f}} \frac {A_j e^{a_j \mu |\mathbf{r}- \mathbf{r}'|}}{|\mathbf{r}- \mathbf{r}'|}.
\end{equation}
Furthermore, due to the restrictions imposed on $A_j$ and $a_j$, Eq.~\ref{eq:sry} reduces to
\begin{equation}
\label{eq:sryoptimised}
V_\mathrm{SR}(\mathbf{r}-\mathbf{r}^\prime)=\frac {A_0 e^{a_0 \mu |\mathbf{r}- \mathbf{r}'|}}{|\mathbf{r}- \mathbf{r}'|}+2\Re\left[\sum_{j=1}^{N_{f}} \frac {A_j e^{a_j \mu |\mathbf{r}- \mathbf{r}'|}}{|\mathbf{r}- \mathbf{r}'|}\right].
\end{equation}
A convolution with the $V_\mathrm{SR}(\mathbf{r}-\mathbf{r}^\prime)$ reduces to a sum of $N_\mathbf{f}+1$ convolutions with $V_\mathrm{Y}(\mathbf{r}-\mathbf{r}^\prime)$ that has the separable form shown above.

\begin{figure}[b]
\includegraphics [width=240pt]{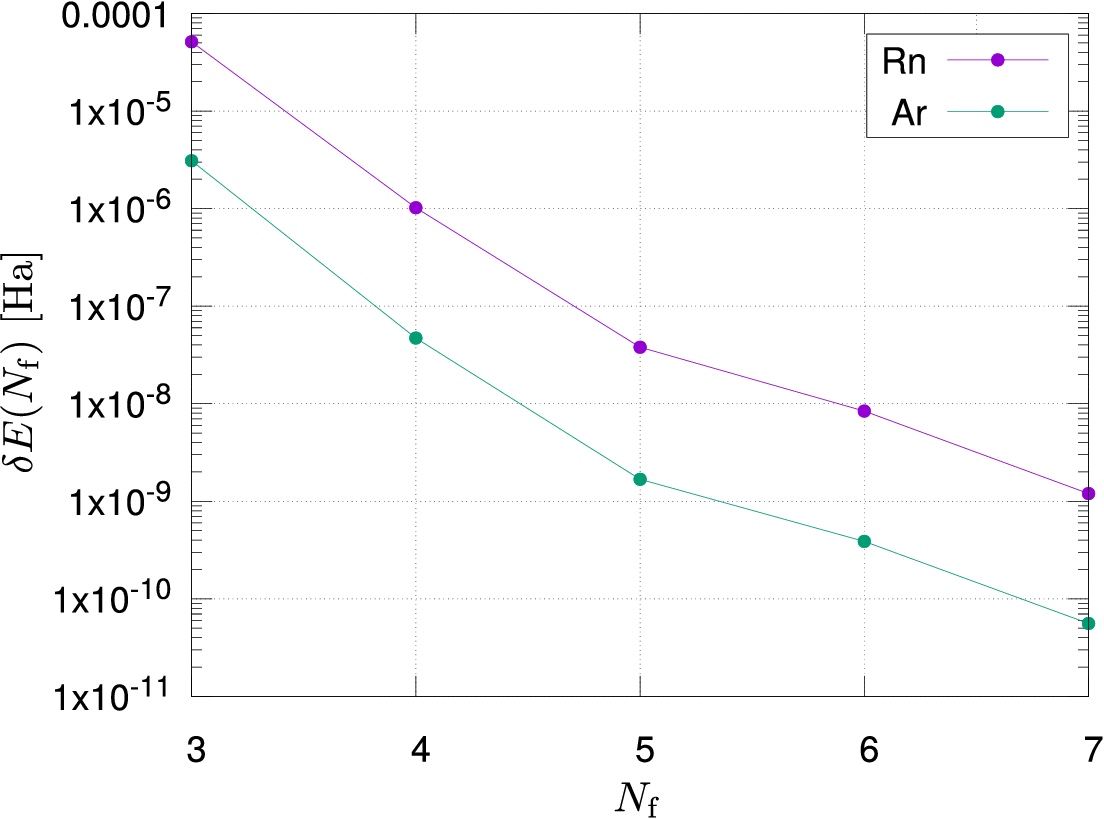}
\caption{\label{fig:LC_BLYP} Errors in LC-BLYP total energies due to the fitted erfc kernel in the non-local exchange. 
The fit employs $2N_\mathbf{f}+1$ functions. 
The energy reference was obtained with $N_\mathrm{f}=8$.}
\end{figure}

To test the introduced approach for the range-separated kernels, we perform a total-energy calculation of the Ar and Rn atoms using the LC-BLYP exchange-correlation functional \cite{Yoshihiro2004}. 
Its non-local part of the exchange consists entirely of the LR term.
Fig.~\ref{fig:LC_BLYP} shows the convergence of the total energy with respect to $N_\mathrm{f}$.
The errors reduce consistently with an increase of the number of fitting functions, and $N_\mathrm{f}=5$ is sufficient to guarantee the precision of $10^{-7}$ Ha for both considered atoms.

\section{Numerical integration and derivatives \label{sec:integrals}}

All radial functions are represented on a predefined grid $r_i$ that spans from $r_\mathrm{min}$ to $r_\mathrm{max}$, and the integrals that appear in this method require a numerical evaluation. 
First, we consider an integral of the kind $\int_{r_{i}}^{r_{i+1}} f(r) dr$.
Assuming $f(r)$ is a smooth function, we interpolate it using a $(2k-1)$th-degree Lagrange polynomial 
\begin{equation}
\label{eq:interp}
L(r)=\sum\limits_{j=i-k+1}^{i+k} f(r_j)\ell_j(r),
\end{equation}
where 
\begin{equation}
\label{eq:lbasis}
\ell_j(r)=\prod_{\substack{j^\prime \ne j}} \frac{r-r_{j^\prime}}{r_j-r_{j^\prime}}.
\end{equation}
$L(r)$ is chosen such that $L(r_j)=f(r_j)$ for $j=i-k+1,\dots,i+k$. The required integral is approximated then as $\int_{r_{i}}^{r_{i+1}} L(r) dr$.
Evaluating the coefficients of this polynomial to compute the integral is impractical.
Instead, we apply a Newton-Cotes formula with $Q+1$ points for numerical integration using equally spaced abscissas \cite{abramowitz1965handbook} and obtain
\begin{equation}
\label{eq:bode}
\int\limits_{r_i}^{r_{i+1}} L(r) dr = \frac{\delta r_i}{Q}\sum\limits_{q=0}^{Q}
W_q L\left(r_i + \frac{q}{Q}\delta r_i \right),
\end{equation}
where $W_q$ are the integration weights and $\delta r_i= r_{i+1}-r_i$.
These integrals can be evaluated using Gaussian quadratures instead, but we choose the Newton-Cotes formulas due to their simplicity, as both approaches yield an exact answer for a polynomial of a sufficiently low degree.
Combining Eqs.~\ref{eq:interp} and \ref{eq:bode}  yields
\begin{equation}
\label{eq:weights1}
\int\limits_{r_i}^{r_{i+1}} L(r) dr = \frac{\delta r_i}{Q}\sum\limits_{j=i-k+1}^{i+k}\sum\limits_{q=0}^{Q}
W_q \ell_j\left(r_i + \frac{q}{Q}\delta r_i \right) f(r_j).
\end{equation}
The result can be written in the compact form
\begin{equation}
  \int\limits_{r_i}^{r_{i+1}} f(r) dr \approx  \sum\limits_{j=i-k+1}^{i+k} w_{ij} f(r_j),
\end{equation}
where 
\begin{equation}
\label{eq:weights}
w_{ij} = \frac{\delta r_i}{Q}\sum\limits_{q=0}^{Q}
W_q \ell_j\left(r_i + \frac{q}{Q}\delta r_i \right) .
\end{equation}
The weights $w_{ij}$ depend only on a grid and have to be computed only once.
If the integration boundaries are $r_\mathrm{min}$ and $r_\mathrm{max}$, the integral can be expressed in a similar form
\begin{equation}
\int\limits_{r_\mathrm{min}}^{r_\mathrm{max}}f(r)dr \approx \sum\limits_{i} \tilde{w}_{i}f(r_i).
\end{equation}

In our further calculations, we use the Bode's rule in Eq.~\ref{eq:bode} and the Lagrange polynomial with $p=9$.
For $i=1,\dots,4$, it is not possible to select the support points for interpolation the same way as suggested above. 
In this case, we interpolate $f(r)$ using $j=0,\dots,9$ assuming that $f(r_0)=0$, and it allows us also to include an integral over the range $0<r<r_1$ in the calculation.
In comparison to neglecting this tiny region, taking it into account makes it possible to define $r_1$ by 2-3 orders of magnitude larger.

Calculation of the kinetic energy requires a derivative, and we use the Lagrange interpolation again. An approximation to the derivative is expressed as 
\begin{equation}
f'(r)=\sum_{j=0}^k  f_j \ell^{(1)}_j(r)
\end{equation}
with the weights
\begin{equation}
\ell^{(1)}_j(r)=\sum_{\substack{i=0 \\ i\ne j }}^k \Biggl( \frac{1}{r_j-r_i}\prod_{\substack{m=0 \\ m\ne j \\ m \ne i}}^k \frac{r-r_m}{r_j-r_m}\Biggr).
\end{equation}

\section{\label{sec:optimal} Radial grids}

\begin{figure}
\includegraphics [width=240pt]{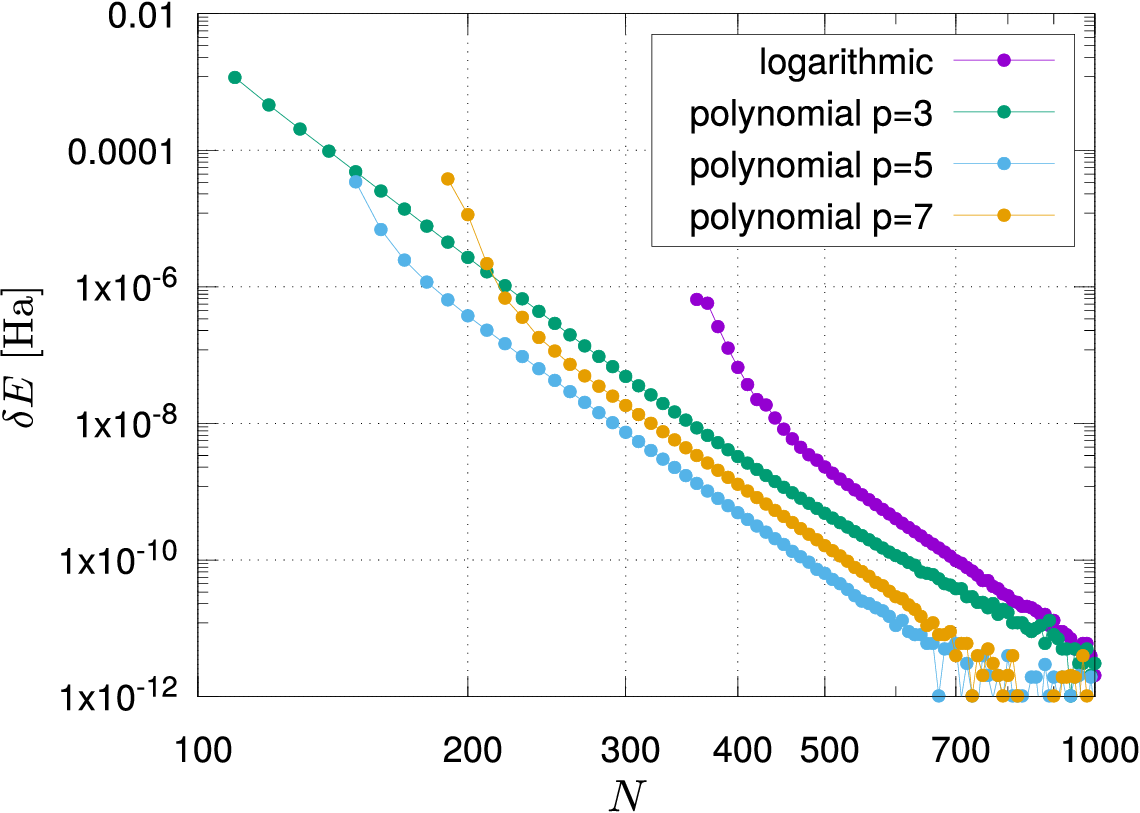}
\caption{\label{fig:E_Ngrid_Ar_HF} 
Error in total HF energy of Ar calculated using a radial grid with $N$ points. 
The reference energy corresponds to a calculation with $N=2000$.
The grid types shown in the legend are defined in Eqs.~\ref{eq:poly} and \ref{eq:log}.}
\end{figure}

\begin{figure}
\includegraphics [width=240pt]{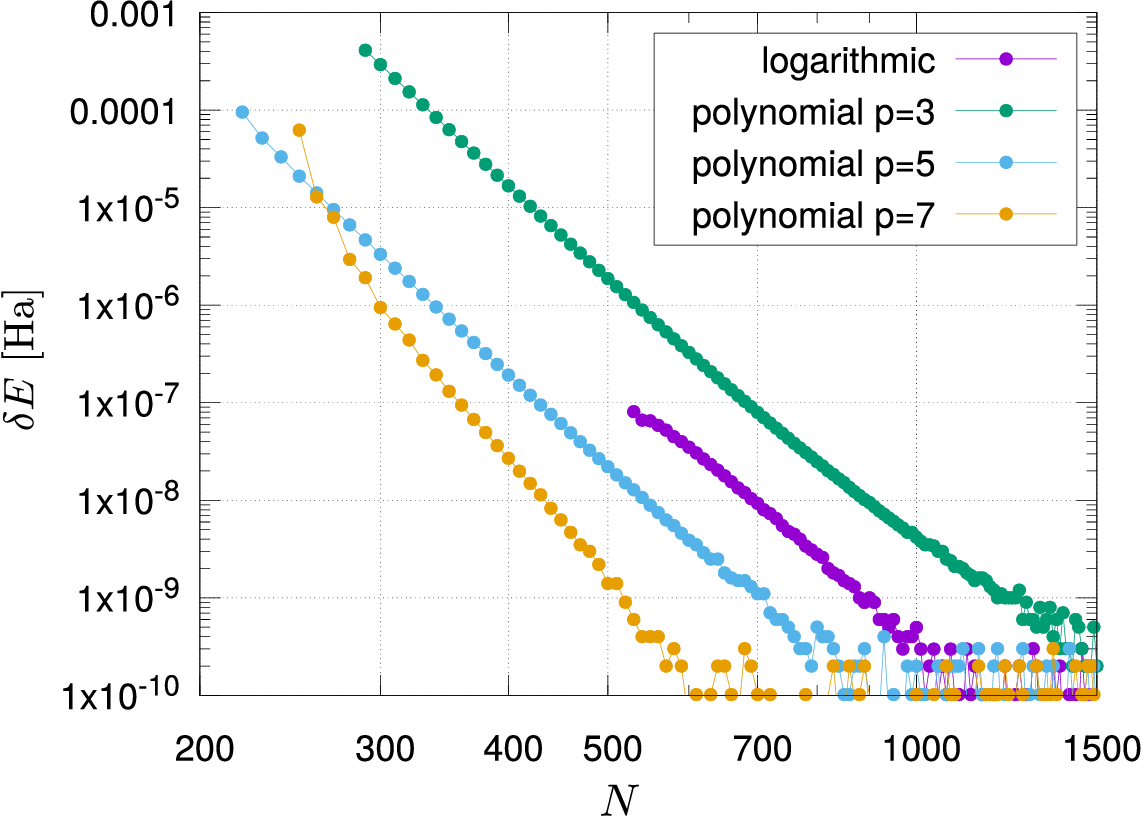}
\caption{\label{fig:E_Ngrid_Rn_HF}
Error in total HF energy of Rn calculated using a radial grid with $N$ points. 
The reference energy corresponds to a calculation with $N=2000$.
The grid types shown in the legend are defined in Eqs.~\ref{eq:poly} and \ref{eq:log}. 
}
\end{figure}

The logarithmic grid is the \textit{de facto} standard choice electronic-structure codes.
It is defined as
\begin{equation}
 r_i=r_\mathrm{min}\left(\frac{r_\mathrm{max}}{r_\mathrm{min}}\right)^{\frac{i-1}{N-1}},
\end{equation}
where $r_\mathrm{min}$ and $r_\mathrm{max}$ correspond to the innermost and the outermost points of the grid, respectively.
$N$ is the number of the grid points.
It was argued in Ref.~\cite{Havlov1984} that an inverse cubic grid is more appropriate in atomic calculations than the equidistant or logarithmic ones.
This idea was tested in the LAPW \texttt{exciting} code \cite{Gulans2014} where the inverse cubic grid is currently the default option.

In this work, we introduce order-$p$ polynomial grids with the following definition:
\begin{equation}
\label{eq:poly}
r_i=ir_\mathrm{min}+\left(\frac{i-1}{N-1} \right)^p (r_\mathrm{max} - Nr_\mathrm{min}).
\end{equation}
The linear term $ir_\mathrm{min}$ ensures that the distance between $r_1$ and $r_2$ is reasonable and roughly is equal to $r_\mathrm{min}$.
Too small $r_2-r_1$ leads to a numerical noise that appears due subtraction of nearly equal numbers during the Lagrange interpolation.
We use the same approach to adjust the logarithmic grid and obtain 
\begin{equation}
\label{eq:log}
 r_i=(i-1)r_\mathrm{min}+r_\mathrm{min}\left(\frac{r_\mathrm{max}}{r_\mathrm{min}}N-1\right)^{\frac{i-1}{N-1}}.
\end{equation}


To assess the described grid types, we compare their performance.  Figs.~\ref{fig:E_Ngrid_Ar_HF} and \ref{fig:E_Ngrid_Rn_HF} show errors in the total energies of the Ar and Rn atoms with respect to $N$.
The inverse polynomial grids with $p=3, 5$ and $7$, perform similarly for Ar, but there is a much larger distinction between them for Rn where the setting $p=7$ performs the best.
In neither case, the logarithmic grid shows the optimal performance. 

\section{Atomic energies \label{sec:results}}

\begin{table*}
\caption{Non-relativistic total energies (in Hartrees) of closed-shell atoms. 
\label{tab:results}}
\setlength{\tabcolsep}{0.10cm} 
\begin{tabular}{lrrrrrr}
\hline
atom & HF & VWN & PBE & PBE0 & B3LYP & LC-BLYP \\ 
\hline
He & -2.861679996 & -2.834835624 & -2.892934867 & -2.895178376 & -2.915218663 & -2.866810561 \\ 
Be & -14.573023168 & -14.447209474 & -14.629947716 & -14.636641425 & -14.673328176 & -14.584722714 \\ 
Ne & -128.547098109 & -128.233481269 & -128.866427745 & -128.871759474 & -128.980973238 & -128.816627071 \\ 
Mg & -199.614636425 & -199.139406315 & -199.955115169 & -199.970695270 & -200.103549936 & -199.907035649 \\ 
Ar & -526.817512803 & -525.946194919 & -527.346128774 & -527.388217197 & -527.567834997 & -527.321256048 \\ 
Ca & -676.758185925 & -675.742282614 & -677.348819102 & -677.392363604 & -677.595272840 & -677.329114053 \\ 
Zn & -1777.848116191 & -1776.573849681 & -1779.182796711 & -1779.191450269 & -1779.503834064 & -1779.205663017 \\ 
Kr & -2752.054977346 & -2750.147940421 & -2753.416108936 & -2753.512330119 & -2753.851525806 & -2753.494137871 \\ 
Sr & -3131.545686439 & -3129.453161377 & -3132.948659228 & -3133.055470684 & -3133.414387383 & -3133.036729104 \\ 
Pd & -4937.921024070 & -4935.368405699 & -4939.793447386 & -4939.923012608 & -4940.317844921 & -4939.893211695 \\ 
Cd & -5465.133142530 & -5462.390982009 & -5467.051582940 & -5467.198445049 & -5467.607200062 & -5467.162476372 \\ 
Xe & -7232.138363872 & -7228.856106486 & -7234.233211984 & -7234.433366465 & -7234.867433857 & -7234.417540215 \\ 
Ba & -7883.543827330 & -7880.111578015 & -7885.731166873 & -7885.930820635 & -7886.384144735 & -7885.925529978 \\ 
Yb & -13391.456193118 & -13388.048594318 & -13395.288895142 & -13395.345840849 & -13395.890072201 & -13395.505337325 \\ 
Hg & -18408.991494945 & -18404.274219990 & -18412.743896725 & -18413.020005924 & -18413.551671176 & -18412.999474421 \\ 
Rn & -21866.772240873 & -21861.346868935 & -21870.576884534 & -21870.947943763 & -21871.481295618 & -21870.907206607 \\ 
Ra & -23094.303666425 & -23088.688083054 & -23098.174709109 & -23098.554874219 & -23099.101083655 & -23098.513007616 \\ 
\hline
\end{tabular}
\end{table*}

We apply the radial solver for calculating non-relativistic total energies of atoms using the following approximations for exchange and correlation: (i) the HF approximation, (ii) the LDA parametrization by Vosko, Wilk and Nusair (VWN5) \cite{Vosko1980}, (iii) the GGA parametrization by Perdew, Burke and Ernzerhof (PBE) \cite{Perdew1996}, (iv) the PBE0 hybrid \cite{Adamo1999}, (v) the B3LYP hybrid \cite{Stephens1994} and (vi) the LC-BLYP hybrid with the long-range part ($\mu=0.3$) of the exchange \cite{Yoshihiro2004}.
The same grid parameters are chosen for all atoms in all calculations, namely, $r_\mathrm{min}=10^{-6}$~bohr, $r_\mathrm{max}=30$~bohr $N=800$.
Based on the findings shown in Sec.~\ref{sec:optimal}, we employ the inverse polynomial grid with $p=7$. 

The total energies obtained for atoms with closed subshells are given in Tab.~\ref{tab:results}.
The complete set of inputs and outputs (total energies, Kohn-Sham eigenenergies and wavefunctions) from our non-relativistic spin-restricted spherically-symmetric calculations for all atoms from H through U is available in an open-access repository~\cite{Uzulis2022data}.
With the chosen parameters, our HF data agree with the results published in Ref.~\cite{Cinal2020} to all decimal places given in Tab.~\ref{tab:results}.
Moreover, for all these elements, we find a 14-digit agreement approaching the double-precision limit in the floating-point representation.

Our VWN5 energies agree perfectly with the data in Ref.~\cite{Lehtola2020}, where these calculations were performed using a finite-element basis.
The total energies were given with signs up to 1 $\mu$Ha, and our results are consistent with all given signs in that work. 
In comparison to another study \cite{Kraisler2010} with VWN5 calculations, we find discrepancies up to 3~$\mu$Ha in the case of the Rn atom.

The agreement between our HF energies and those in Ref.~\cite{Cinal2020} as well as the convergence tests performed in Sec.~\ref{sec:optimal} make us confident that all digits given in the VWN5 results are significant.
The remaining four exchange-correlation functionals employ the GGA, and, in this case, it is common to introduce density thresholds below which $v_\mathrm{xc}^\mathrm{L}(\mathbf{r})$ is not computed. 
Such a threshold is also introduced in the \texttt{libxc} library and is the main precision-limiting factor in these calculations.
Nevertheless, we anticipate that this feature does not lead to errors exceeding a few nHa. 

Our PBE and PBE0 energies for He, Be, Ne, Mg and Ar agree with the multi-resolution analysis data well within 1 $\mu$Ha. 
Finally, also the LC-BLYP total energies for He, Be, Ne and Mg agree perfectly (within all signs given in the reference) with the data from the finite-element calculation reported in Ref.~\cite{Lehtola2020}, and only for Ar with find a difference of 1~$\mu$Ha.
The obtained agreement in the LC-BLYP data is noteworthy, because the two methods employed different approaches for calculating the integrals with the long-range part of the exchange.

\section{Conclusions \label{sec:conclusions}}
We implemented a highly precise tool for solving the Kohn-Sham problem for spherically-symmetric atoms.
This tool solves the Schr\"{o}dinger equation as an integral equation rather than a differential one and heavily relies one-dimensional convolutions.
For this reason, we implemented efficient quadratures based on Lagrange interpolation and investigated different grid types.
The frequently used logarithmic grids are far from being the optimal choice.
The proposed alternative, inverse polynomial grids, are superior to the logarithmic ones (especially with the polynomial degree 7) and allow for a significant reduction of the number of points.
These technical tricks are potentially useful in electronic-structure codes, especially, those employing linearized augmented plane waves or linearized muffin-tin orbitals. 

To support hybrid functionals with the screened or long-range exchange in the presented radial solver, we introduced a new method for calculating convolutions with the complementary error function kernel. 
This method presents an alternative to the standard approach that applies bi-variate quadrature. 
Our approach can be transferred to full-potential all-electron codes where it will serve as a reference method for hybrid functionals with screened exchange.

Finally, we applied the radial solver in calculations of closed-shell atoms and observed a remarkable agreement with highly precise Hartree-Fock data in literature.
This result implies that our further calculations performed with five local and hybrid exchange-correlation functionals set a reliable benchmark.  

\begin{acknowledgments}
This work was funded by the Latvian Research Council via the project Precise Fock Exchange (PREFEX) with the grant agreement No. lzp-2020/2-0251.
\end{acknowledgments}

\nocite{*}

\providecommand{\noopsort}[1]{}\providecommand{\singleletter}[1]{#1}%

\end{document}